\documentclass[10pt,conference]{IEEEtran}
\IEEEoverridecommandlockouts

\usepackage{amsmath,amssymb,amsfonts}
\usepackage{algorithmic}
\usepackage{array}
\usepackage{graphicx}
\usepackage{comment}
\usepackage{xurl}
\usepackage{float}
\usepackage[hidelinks]{hyperref}
\usepackage{textcomp}
\usepackage{xcolor}
\usepackage{french}
\usepackage{pbalance}
\usepackage[backend=biber,style=numeric,sorting=none]{biblatex}
\makeatletter
\newcommand{\linebreakand}{%
  \end{@IEEEauthorhalign}
  \hfill\mbox{}\par
  \mbox{}\hfill\begin{@IEEEauthorhalign}
}
\makeatother

\begin{document}
\title{From Conceptual Scaffold to Prototype: A Standardized Zonal Architecture for Wi-Fi Security Training\\
}

\author{%
\IEEEauthorblockN{Vyron Kampourakis\IEEEauthorrefmark{1},
Efstratios Chatzoglou\IEEEauthorrefmark{2},
Vasileios Gkioulos\IEEEauthorrefmark{1},
Sokratis Katsikas\IEEEauthorrefmark{1}}
\IEEEauthorblockA{\IEEEauthorrefmark{1}\textit{Dept. of Information Security and Communication Technology}\\
\textit{Norwegian University of Science and Technology}\\
2802 Gjøvik\\
\{vyron.kampourakis, vasileios.gkioulos, sokratis.katsikas\}@ntnu.no}
\IEEEauthorblockA{\IEEEauthorrefmark{2}\textit{Dept. of Information and Communication Systems Engineering}\\
\textit{University of the Aegean}\\
83100 Greece\\
efchatzoglou@aegean.gr}}

\maketitle

\begin{abstract}
Wi-Fi is the dominant wireless access technology, but its widespread use also exposes systems to threats such as rogue access points, deauthentication attacks, and other IEEE 802.11-specific vulnerabilities. Although Cyber Ranges (CRs) have become valuable platforms for cybersecurity training and experimentation, existing wireless-oriented solutions mainly target heterogeneous IoT or mobile-network settings, with Wi-Fi typically treated as one among many. As a result, dedicated CR environments for Wi-Fi-specific security experimentation remain limited. This gap is particularly relevant because wireless attacks often require protocol-aware experimentation that is difficult to reproduce in conventional training environments. This paper introduces a conceptual architecture for a Wi-Fi-focused CR tailored to IEEE 802.11 security scenarios and an open-source prototype. The proposed design is grounded in established CR design principles and organized around core infrastructure, learning management and support, monitoring, management, and access-control zones. Structuring the platform into these distinct zones, the architecture supports modularity, scalability, and future extensibility. Part of the design is realized in a prototype publicly available in a GitHub repository that implements the scenario generation, storage, retrieval, and instantiation workflow, offering an initial practical foundation for the proposed architecture. Overall, the paper provides a structured foundation for the future implementation of Wi-Fi-specialized CR platforms for targeted experimentation.
\end{abstract}


\section{Introduction}
\label{S:intro}

The predominant protocol for Wireless Local Area Networks (WLANs) is IEEE 802.11, universally known as Wi-Fi, which is a fundamental enabler of modern digital connectivity, supporting a vast range of personal, enterprise, and, in recent years, even mission-critical applications~\cite{kampourakis2025numeris}. The increasing demand for mobility, flexibility, and cost-effective network deployment has driven its widespread adoption. At the same time, the rapid evolution of connected devices such as Internet of Things (IoT) ecosystems has made wireless communication more pervasive than ever. Importantly, the advent of Wi-Fi 6 (802.11ax) and the emerging Wi-Fi 7 (802.11be) have introduced significant advancements in throughput, latency, security, and overall performance, making them well-suited for environments where reliable and high-performance wireless connectivity is essential. However, the growing reliance on advanced wireless technologies also introduces new attack vectors, including over-the-air exploits, rogue access points, unauthorized associations, and protocol-specific vulnerabilities. In practice, Wi-Fi networks can serve as attractive entry points for adversaries, potentially enabling them to bypass traditional perimeter defenses and compromise connected systems.

Despite the significance of these threats, educating the future cybersecurity workforce in wireless security remains an underdeveloped and often underfunded~\cite{Berhanu2026} area, with most cybersecurity education methods still relying on traditional approaches such as lectures and seminars~\cite{vykopal2017lessons}. It is not a clich\`e that humans are the weakest link in the cybersecurity chain~\cite{human:factor}; therefore, particular attention should be devoted to improving cybersecurity knowledge, skills, and awareness at all levels. To this end, Cyber Ranges (CRs) have emerged as a fitting solution for diverse audiences, offering a versatile approach to strengthening cybersecurity posture~\cite{vkamp:2025}. These platforms have attracted notable interest from both the scientific community and the private sector due to their dual role. First, they can be used to raise awareness and educate personnel about cybersecurity best practices. Second, they can serve as valuable platforms for uncovering zero-day vulnerabilities and evaluating suitable security controls for the system under assessment.

The challenges outlined above reveal a dual gap in current practices. On one hand, education and training efforts often fail to adopt practical, scenario-driven environments that reflect the complexity of modern wireless ecosystems. On the other hand, CRs that explicitly address the security implications of contemporary Wi-Fi technologies, particularly IEEE 802.11, remain limited in the literature, despite the growing ubiquity of wireless connectivity. This lack of focused training environments leaves organizations and practitioners insufficiently prepared to detect, contain, and mitigate Wi-Fi–borne threats. To address this gap, we propose the conceptual architecture of a CR platform focused on Wi-Fi environments, capable of realistically emulating wireless infrastructures and simulating protocol-specific attack scenarios.

The rest of the paper is structured as follows. Section~\ref{S:Rel:work} reviews related work on CR and wireless testbeds, highlighting the gap in Wi‑Fi‑focused security platforms. Section~\ref{S:arch} presents the proposed Wi‑Fi‑focused CR architecture, detailing its structural view. Section~\ref{SS:otherviews} describes excerpts of the functional and informational views, with a focus on the implemented scenario generation and instantiation workflow. Section~\ref{S:lim:fut} discusses the main limitations of the proposed architecture and outlines planned extensions and future work. The last section concludes the paper.

\section{Related Work}
\label{S:Rel:work}

Several CRs and large-scale testbeds have been proposed for IoT, wireless, and mobile-network systems. However, most of these platforms are better characterized as testbeds or hybrid emulation environments, since they primarily support experimentation with heterogeneous physical and virtual components rather than a systematically organized cyber range architecture. In this sense, it is important to distinguish testbeds from cyber-(physical) ranges: whereas testbeds emphasize controlled experimentation, cyber ranges aim to provide a more integrated environment for scenario-driven training and assessment~\cite{kava2019}. In this context, most focus on heterogeneous environments combining physical and virtual components, supporting protocols such as ZigBee, BLE, Wi-Fi, IEEE 802.15.4, and cellular/5G. Representative examples include hybrid IoT CRs integrating emulated and real devices~\cite{balto_IoT_2023}, IoT training platforms based on physical sensor networks and simulators~\cite{IoT_CR_2020}, smart-home and CPS-oriented CRs~\cite{IoT_virt_2017}, as well as 5G- and digital twin-based environments~\cite{NITRO_2024}. Some works also propose conceptual architectures for CR delivery as a service~\cite{CyberIoT_2021}.

Despite this diversity, three limitations emerge. First, most platforms rely on ad hoc testbeds or custom architectural designs rather than standardized CR reference models~\cite{balto_IoT_2023, IoT_virt_2017, NITRO_2024, ABUWARAGA_2020}. Second, evaluation is typically demonstration-based or requirement-driven, often using non-uniform metrics that hinder reproducibility and cross-platform comparison~\cite{IoT_CR_2020, IoT_virt_2017, CyberIoT_2021}. Third, although several testbeds include Wi-Fi among the supported technologies, none is specifically designed around Wi-Fi cybersecurity as its primary focus~\cite{balto_IoT_2023, ABUWARAGA_2020}. In most cases, Wi-Fi appears only as one protocol among many, without dedicated support for IEEE 802.11-specific threats and experimentation. This gap is significant because Wi-Fi remains the dominant WLAN technology and exposes a distinct attack surface, including rogue access points, deauthentication attacks, and protocol-specific weaknesses~\cite{deauth_equals_2021, krack, Dragonblood, vanhoef2021fragment}. Therefore, there is a clear need for a CR tailored to Wi-Fi environments that allows for protocol-aware experimentation, training, and evaluation under realistic conditions. In contrast to prior work, our study introduces a CR architecture explicitly dedicated to Wi-Fi, with emphasis on IEEE 802.11-specific scenarios and a systematic architectural perspective.

\section{Wi-Fi Cyber Range Architecture}
\label{S:arch}

The proposed architecture builds on the CR reference model of~\cite{vkamp:2025}, which aligns with the ISO/IEC/IEEE 42010 standard and supports the description of CRs through structural, functional, and informational views. We adopt this model because it provides a systematic basis for organizing CR components, grounded in National Institute of Standards and Technology (NIST) and European Cyber Security Organisation (ECSO) guidelines and informed by the state-of-the-art literature. Moreover, similar reference architectures proposed in the literature~\cite{kava2019,Ukwandu2020,YAMIN_CR_2020,Katsantonis2023} operate at a high level of abstraction, primarily focusing on the operational aspects of CRs, such as human involvement and internal data flows, while largely neglecting critical elements such as development processes, structural design, and technical implementation, including the underlying technologies and the interactions both within and across architectural layers. 

\begin{figure*}[!htbp]
    \centering
    \includegraphics[width=0.9\linewidth]{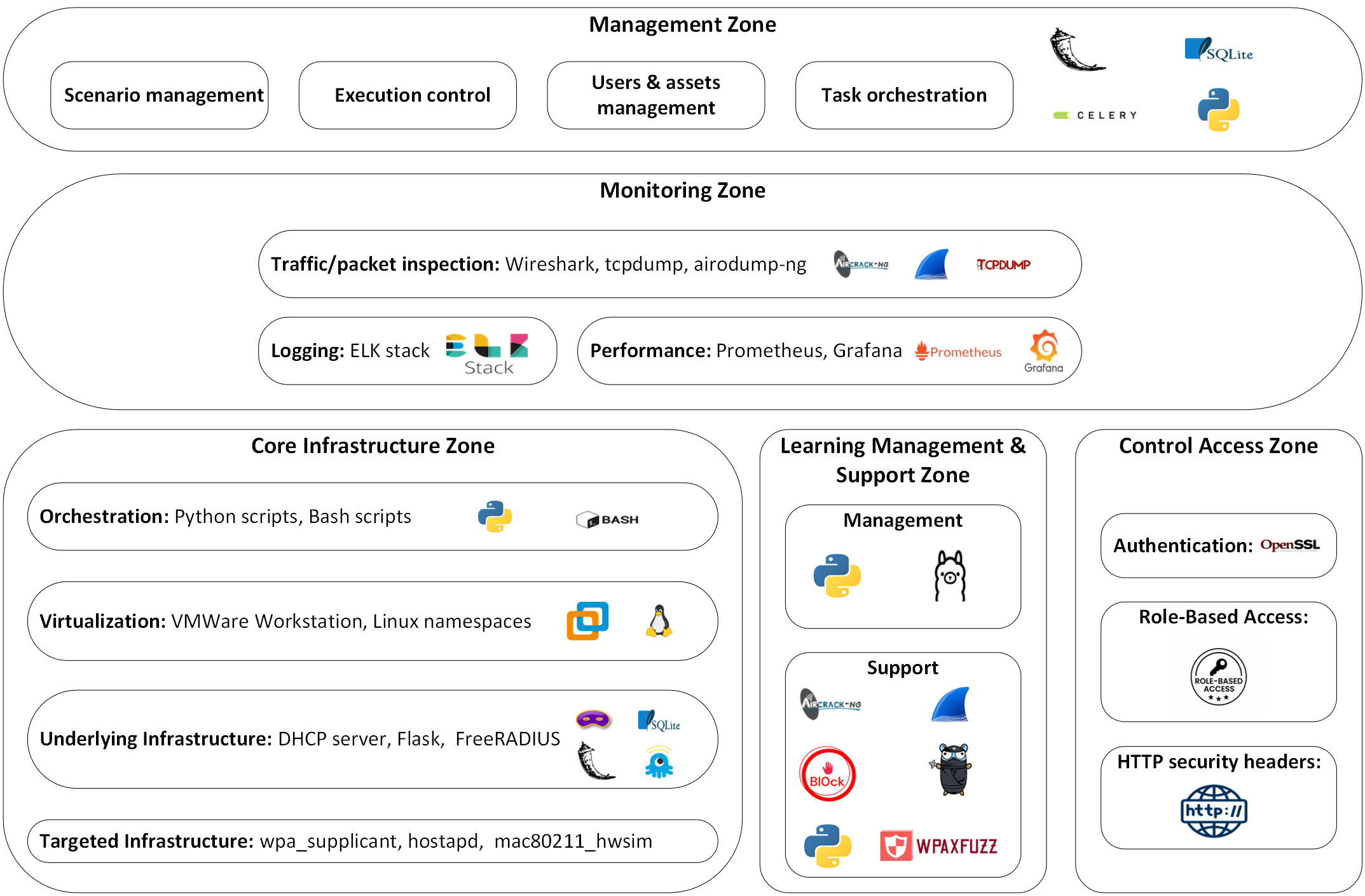}
    \caption{Structural view of the proposed Wi-Fi CR.}
    \label{F:structural:view}
\end{figure*}

\sloppy{Additionally, these works tend to loosely adapt guidelines from widely recognized standards and demonstrate only partial alignment with them, often employing varying terminology to describe similar concepts; thus, they introduce some ambiguity and reduce uniformity, potentially limiting the effectiveness of standardization efforts. Furthermore, the modular organization followed in~\cite{vkamp:2025} facilitates future adaptation to new IEEE 802.11 amendments and emerging features by allowing updates at the orchestration, scenario-definition, and targeted-infrastructure layers without redesigning the full platform. Altogether, the reference architecture proposed in~\cite{vkamp:2025} serves as the conceptual scaffold for the proposed Wi-Fi-focused CR. Figure~\ref{F:structural:view} presents the structural view of the architecture. At a high level, the platform comprises five main zones: i) core infrastructure, ii) Learning Management \& Support, iii) monitoring, iv) management, and v) control access. Together, these support the definition, deployment, supervision, and controlled use of Wi-Fi-specific CR scenarios.}

\subsection{Core Infrastructure}
\label{SS:core}

The core infrastructure zone serves as the operational backbone of the proposed Wi-Fi CR, providing the mechanisms to instantiate, isolate, and manage scenario-specific testbeds. As shown in Figure~\ref{F:structural:view}, it is organized into four layers: orchestration, virtualization, underlying, and targeted infrastructures. The orchestration layer coordinates the deployment and lifecycle of scenarios. It relies on Python-based control logic and Bash scripts to automate topology creation, service initialization, and scenario-specific execution workflows. The virtualization layer provides lightweight isolation through Linux-based mechanisms. Specifically, \texttt{Linux namespaces} are used to create independent environments for emulated access points (APs) and stations (STAs), enabling multiple wireless nodes to coexist on a single virtual host (\texttt{VMWare}) while remaining logically separated. 

The underlying infrastructure layer comprises the components required for scenario execution, including the Flask-based web interface, SQLite databases, and auxiliary services for networking. Together, these components enable user interaction, persistent scenario storage, and coordination with the orchestration mechanisms. The targeted infrastructure layer corresponds to the Wi-Fi testbed instantiated during scenario execution. It is built on the \texttt{mac80211\_hwsim} kernel module for software-based IEEE 802.11 radio emulation, combined with user-space services such as \texttt{hostapd} for AP and \texttt{wpa\_supplicant} for STA instantiation, respectively. In this way, the platform can reproduce realistic wireless environments with multiple Basic Service Sets (BSSs), potentially also forming Extended Service Sets (ESSs), spanning from Small Office/Home Office (SOHO) to 802.1X/EAP-enabled configurations and security setups, providing a flexible and reproducible basis for Wi-Fi-specific experimentation. Altogether, this block provides the software-defined Wi-Fi environment in which scenarios are instantiated. Its main role is to emulate realistic IEEE 802.11 network topologies while remaining lightweight, reproducible, and isolated from production wireless infrastructures.

\subsection{Learning Management \& Support Zone}
\label{SS:LMS}

As depicted in Figure~\ref{F:structural:view}, the learning management and support zone provides the modules through which instructors can define scenarios and learners interact with the CR in a practice-oriented manner. In the proposed Wi-Fi CR, this zone bridges educational requirements with the core infrastructure, ensuring that scenarios can be created, customized, deployed, and exercised in a reproducible way.

From the learning management perspective, the central module of this zone is scenario generation and development. Instructors define Wi-Fi security scenarios through a \texttt{Flask}-based web interface, either by configuring pre-defined topology blueprints or by providing natural-language descriptions to a local LLM (\texttt{Llama}), which converts these high-level inputs into structured representations suitable for automated deployment. The resulting scenario definitions are stored in the scenario database. In this way, the platform supports both manual and semi-automated scenario creation, enabling instructors to prepare exercises of varying complexity while reducing the burden of low-level configuration. Importantly, this module directly interfaces with the core infrastructure zone, as each scenario determines how the targeted Wi-Fi testbed is instantiated, including the number of APs and STAs, namespace allocation, DHCP provisioning through tools such as \texttt{dnsmasq}, and, where required, 802.1X/EAP authentication through services such as \texttt{FreeRADIUS}.

\sloppy{A second important function of this zone relates to learners' support utilities. Specifically, to enable realistic experimentation, the proposed platform integrates offensive, defensive, and analysis tools that learners can use during exercises, thereby providing them with the practical means to apply specific Tactics, Techniques, and Procedures (TTPs). In the present design, these include offensive utilities from the \texttt{Aircrack-ng} suite, packet capture and inspection tools such as \texttt{tcpdump}, \texttt{tshark}, and \texttt{Wireshark}, as well as Wi-Fi-specific offensive tools such as \texttt{WPAxFuzz}~\cite{wpaxfuzz} and \texttt{Bl0ck}~\cite{Bl0ck} for protocol-aware experimentation. Collectively, these tools allow learners to perform activities such as wireless discovery, traffic inspection, handshake analysis, deauthentication testing, fuzzing of Wi-Fi Protected Access (WPA) implementations, and broader assessment of IEEE 802.11 behavior under both benign and adversarial conditions. In this sense, the learning support component equips the learner with the means required to carry out the tactical objectives of a given exercise.}

In addition, this zone may support auxiliary mechanisms that improve the quality of exercises, such as automated validation routines, predefined task workflows, and extensible scenario logic. For example, post-deployment checks based on \texttt{ping}, ARP resolution, and \texttt{iperf3} can be used to verify that a scenario has been instantiated correctly before learner interaction begins. More broadly, the design allows future integration of richer automation mechanisms, such as user-activity scripting, traffic generation, or AI-assisted guidance, to increase realism and reduce repetitive setup effort. Overall, the learning management and support zone ensures that the proposed Wi-Fi CR is not merely an emulated infrastructure, but a usable and educationally meaningful environment in which scenarios can be systematically authored, deployed, and exercised.

\subsection{Monitoring Zone}
\label{SS:Mon:zone}

\sloppy{The monitoring zone provides visibility into the behavior of both the instantiated Wi-Fi infrastructure and the users interacting with it, enabling analysis, debugging, and evaluation of training scenarios. As illustrated in Figure~\ref{F:structural:view}, it integrates tools for packet-level inspection, centralized logging, and performance monitoring. At the network level, traffic inspection is supported through tools such as \texttt{tcpdump}, \texttt{tshark}, and \texttt{Wireshark}, which allow capturing and analyzing IEEE 802.11 frames generated within the emulated environment. Additionally, Wi-Fi-specific monitoring utilities from the \texttt{Aircrack-ng} suite (\texttt{Airodump-ng}) enable observation of BSS or ESS activity, client associations, and channel usage, facilitating the study of protocol behavior and attack scenarios.}

For system-level monitoring, the platform incorporates performance monitoring tools such as \texttt{Prometheus} and \path{Grafana}, which provide insights into resource utilization (e.g., CPU, memory, and network throughput) during scenario execution. These metrics are particularly useful for assessing the scalability and responsiveness of the CR under varying scenarios and audiences. Furthermore, centralized logging mechanisms (e.g., \texttt{ELK stack}) aggregate logs from core services such as \texttt{hostapd}, \texttt{wpa\_supplicant}, and authentication components, enabling event correlation and post-exercise analysis. Together, these allow both instructors and learners to observe system behavior in real time and retrospectively evaluate the outcomes of training exercises.

\subsection{Management Zone}
\label{SS:Man:zone}

The management zone is responsible for the operational control and lifecycle management of the CR, ensuring coordinated execution of scenarios and efficient use of system resources. As depicted in Figure~\ref{F:structural:view}, it encompasses components for scenario management, execution control, and user/session coordination. At its core, a Flask-based management serves as the control plane of the platform, enabling the creation, modification, and retrieval of scenario definitions, as well as interaction with the underlying infrastructure. Scenario metadata and execution state are maintained in the SQLite database, allowing persistent storage and tracking of scenario instances. The same management layer also handles users, where their information is stored in the SQLite database and can be administered by privileged users (e.g., administrators), supporting operations such as account creation and deletion, password management, and user information retrieval. 

To support controlled and scalable execution, the platform incorporates a task orchestration mechanism (e.g., \texttt{Celery}), which enables asynchronous deployment and management of scenarios. This allows multiple learners to interact with the system concurrently while ensuring that scenario instantiation, execution, and teardown are handled reliably. In addition, resource management functions oversee the allocation and cleanup of system resources, particularly the creation and removal of network namespaces and associated services. This ensures isolation between concurrent scenarios and prevents resource leakage across sessions. User and session management mechanisms further map learners to active scenarios, enabling controlled access and coordinated interaction with the instantiated environments.

\subsection{Control Access Zone}
\label{SS:ctrl:zone}

The control access zone enforces security-by-design principles by regulating authentication and authorization across the CR. Its primary role is to ensure that only authorized users can access specific resources and interact with scenario instances, thereby preserving the integrity and isolation of the platform. Authentication is implemented through standard mechanisms such as credential-based login. Authorization is enforced through role-based access control (RBAC), distinguishing between different user roles such as instructors and learners. This enables fine-grained control over actions, for example, restricting scenario creation and modification to instructors, while allowing learners to browse, select, and execute predefined scenarios. Access policies also govern the interaction with instantiated infrastructures, ensuring that users can only access the environments assigned to them. 

\begin{figure*}[!ht]
    \centering
    \includegraphics[width=0.9\linewidth]{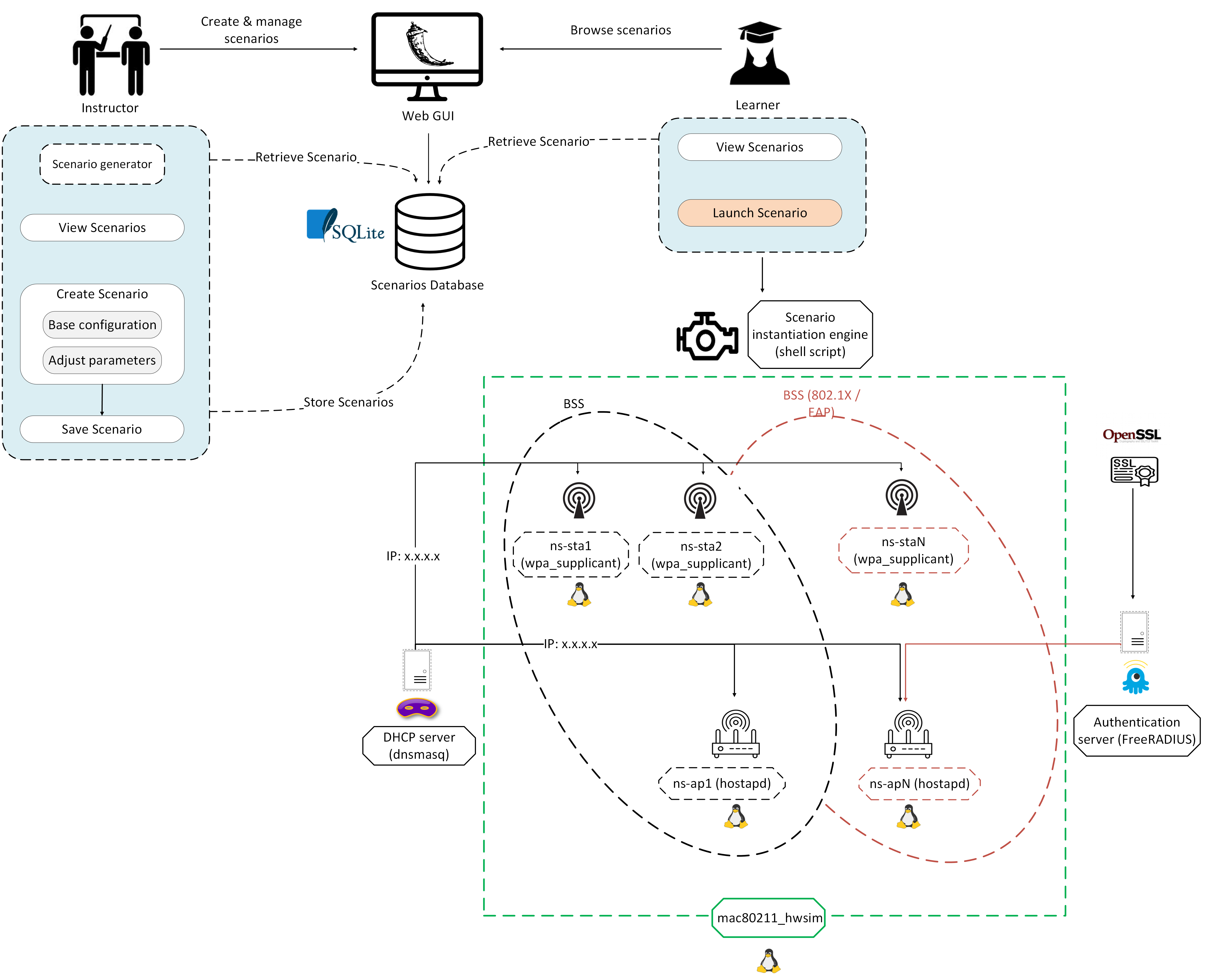}
    \caption{Functional view's excerpt: Scenario generation and instantiation of a namespace-based IEEE 802.11 testbed.}
    \label{F:fun:view}
\end{figure*}

In addition, access control mechanisms contribute to the isolation of concurrent sessions by binding user identities to specific scenario instances. This prevents unauthorized cross-access between learners and maintains the confidentiality of ongoing exercises. To further strengthen the security of the web-facing components, the platform also employs protective Hypertext Transfer Protocol (HTTP) response headers to mitigate common web threats such as Cross-Site Scripting (XSS), Cross-Site Request Forgery (CSRF), and clickjacking, thereby reinforcing the overall security posture of the control access layer.

\section{Functional and Informational Views}
\label{SS:otherviews}

This subsection briefly presents excerpts of the functional and informational views of the proposed Wi-Fi CR, based on the structural view detailed in section~\ref{S:arch}, focusing on the implemented workflow for scenario generation, storage, retrieval, and instantiation. Although the proposed Wi-Fi CR is presented as a conceptual architecture, the components depicted in the functional and informational excerpts correspond to an implemented prototype of the scenario generation and instantiation pipeline. In practice, this orchestration layer acts as the bridge between abstract scenario descriptions and the actual Wi-Fi targeted infrastructure, ensuring that each selected configuration is deployed consistently and with minimal manual intervention. This prototype is publicly available in a GitHub repository~\cite{CRrepo}.

\begin{figure*}[!ht]
    \centering
    \includegraphics[width=0.8\linewidth]{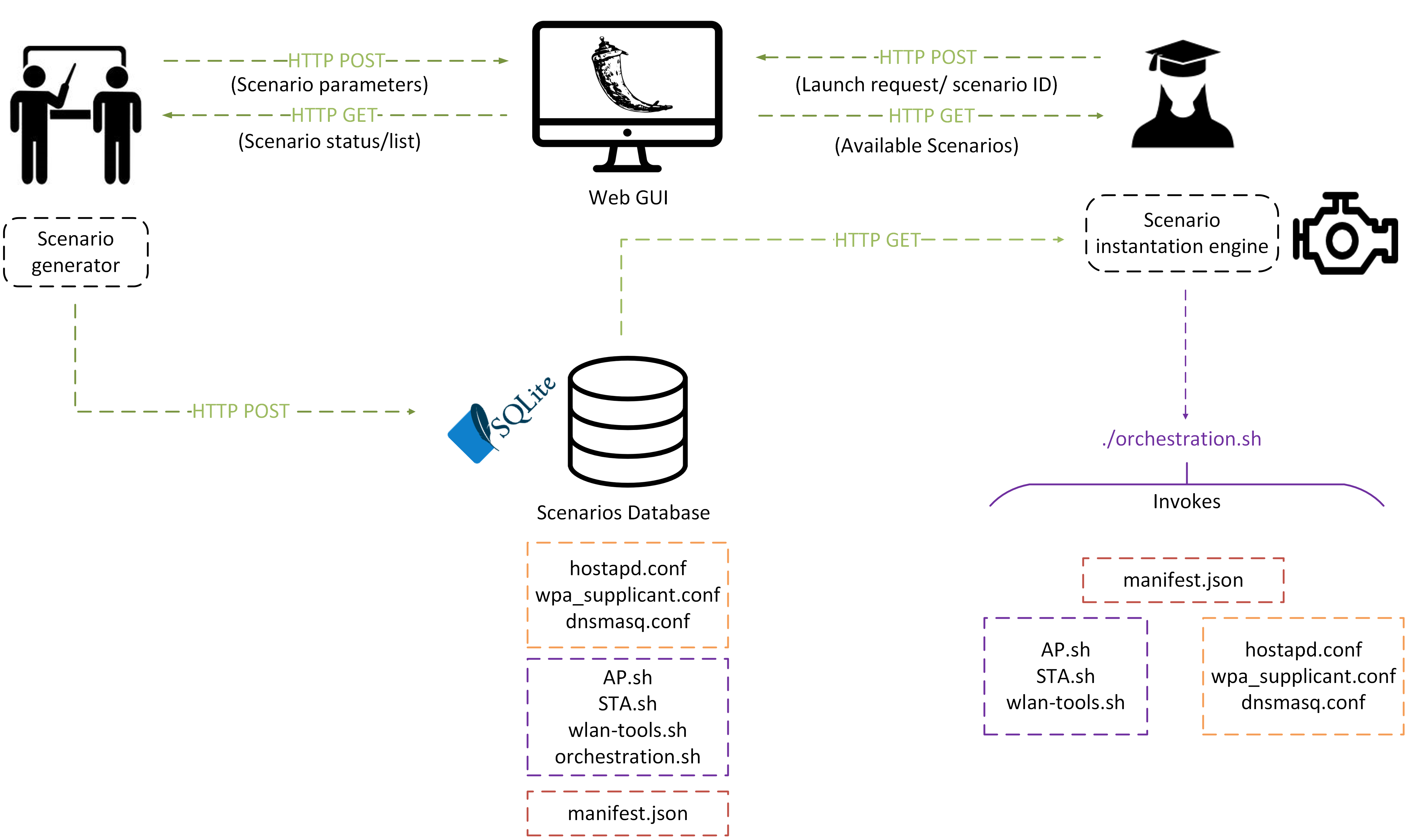}
    \caption{Informational view's excerpt: Scenario generation and instantiation of a namespace-based IEEE 802.11 testbed.}
    \label{F:inf:view}
\end{figure*}

Figure~\ref{F:fun:view} illustrates an excerpt of the functional view, focusing on the workflow from scenario definition to testbed instantiation. Initially, the instructor defines and manages scenarios through a web-based interface, where configurations such as topology and security parameters are specified and stored in a centralized \texttt{SQLite} database. The learner can subsequently browse the available scenarios, select one of interest, and trigger its execution. This action invokes the scenario instantiation engine, implemented as a shell-based orchestration component, which deploys the corresponding targeted infrastructure. In addition to supporting scenario launch, the functional flow also ensures that the selected configuration is validated and mapped to the appropriate deployment scripts before instantiation begins.

\sloppy{The instantiated testbed consists of a namespace-driven Wi-Fi environment built upon the \texttt{mac80211\_hwsim} module~\cite{Fontes2015}, enabling the emulation of diverse topologies with potentially multiple APs and STAs. Depending on the selected scenario, the infrastructure may include both private BSS configurations and 802.1X/EAP-enabled BSSs. Supporting services are provisioned accordingly, including Dynamic Host Configuration Protocol (DHCP) functionality via \texttt{dnsmasq} and certificate-based authentication through \texttt{OpenSSL} and \texttt{FreeRADIUS}. This design allows the same orchestration logic to support both simple training cases and more advanced authentication-focused exercises. Overall, this workflow highlights how high-level scenario definitions are transformed into fully operational, isolated Wi-Fi testbeds~\cite{Rusca2023, Riliskis2015}.}

Figure~\ref{F:inf:view} illustrates an excerpt of the informational view, focusing on the data artifacts and exchanges involved in scenario definition and instantiation. In the proposed platform, instructors interact with the web GUI through HTTP \texttt{POST} requests to submit scenario parameters and through HTTP \texttt{GET} requests to retrieve scenario status or previously defined scenarios. Similarly, learners use HTTP \texttt{GET} requests to browse the available scenarios and HTTP \texttt{POST} requests to submit a launch request for a selected scenario. These interactions are mediated by the backend, which stores and retrieves the corresponding artifacts from the SQLite scenarios database. This improves traceability by preserving the relationship between user input, configuration files, and execution scripts.

\sloppy{Each stored scenario is represented as a collection of structured and executable artifacts. These include service configuration files, such as \texttt{hostapd.conf}, \texttt{wpa\_supplicant.conf}, and \texttt{dnsmasq.conf}, a topology manifest (\texttt{manifest.json}) describing namespace allocation and interface roles, and a set of role-specific and orchestration shell scripts, such as \texttt{AP.sh}, \texttt{STA.sh}, \texttt{wlan-tools.sh}, and \texttt{orchestration.sh}. Taken together, these artifacts separate the user-facing description of a scenario from its low-level execution logic, which improves modularity, reusability, and ease of maintenance. Upon scenario selection, the scenario instantiation engine retrieves the required artifacts and invokes the main orchestration script. This script, in turn, uses the manifest, configuration files, and auxiliary shell scripts to materialize the corresponding namespace-based Wi-Fi testbed. As a result, deployment is not only automated but also reproducible across repeated runs. Therefore, the informational flow captures not only the storage and retrieval of scenario data, but also the dependency chain through which configuration, orchestration, and execution are coordinated. In this way, the informational flow captures how user requests are translated into stored scenario artifacts and how these artifacts are subsequently consumed to instantiate an operational IEEE 802.11 environment.}

\section{Limitations and Future Work}
\label{S:lim:fut}

The proposed Wi‑Fi‑focused CR architecture has several limitations that should be acknowledged. First, the current design and prototype target IEEE 802.11‑based wireless environments and do not natively integrate other wireless technologies such as cellular (e.g., 5G), Bluetooth, or competing WLAN protocols; this restricts its applicability to strictly Wi‑Fi‑centric scenarios. Second, the implemented namespace‑based prototype relies on software‑defined emulation via \texttt{mac80211\_hwsim} and co‑existing user‑space tools, which may not fully capture the behavior of large‑scale, heterogeneous deployments involving real hardware radios, interference, and propagation effects. Third, the scenario generation and orchestration workflows are currently focused on single‑user or small‑cohort experiments, and the platform’s scalability and performance under many concurrent learners or complex multi‑BSS topologies remain to be fully evaluated.

In addition, the architectural treatment of human‑centric aspects, including detailed learning objectives, assessment metrics, and fine‑grained didactic workflows, is still conceptual and not yet supported by a comprehensive evaluation framework. The current implementation covers the core functional and informational flows for scenario generation, storage, retrieval, and instantiation, but it does not yet incorporate advanced monitoring or adaptive guidance mechanisms that could tailor exercises to individual learner profiles or dynamically adjust scenario difficulty based on performance. Moreover, the control access and security policies are formulated at a high level; the integration of standardized security‑by‑design patterns, such as more granular attribute‑based access control or policy‑driven runtime enforcement, remains an open direction.

Given the above limitations, future work will focus on three main directions. First, the open‑source implementation will be extended to cover all architectural components, ensuring that the full Wi‑Fi‑focused CR can be instantiated and exercised in a uniform, reproducible pipeline. In this direction, further efforts will focus on revisiting the representation and assessment perspectives through formal modeling for structured, machine-interpretable CR descriptions~\cite{kamponto2026}, as well as multi‑criteria evaluation frameworks that enable explainable and quantitative assessment~\cite{kampourakis2025llm} of CR capabilities and design trade‑offs. These extensions will help establish a more systematic basis for comparing Wi‑Fi‑focused CRs with other security training environments and will mature the proposed architecture into a fully operational, extensible Wi‑Fi‑specialized CR for targeted IEEE 802.11 security experimentation and training.

\section{Conclusions}
\label{S:conc}

This paper introduced a conceptual architecture for a CR tailored to Wi-Fi environments. In contrast to existing CRs and testbeds, where Wi-Fi is typically treated as one communication technology among many, the proposed design places IEEE 802.11 at the core of the platform and emphasizes protocol-aware scenario generation, reproducible deployment, and structured architectural organization. The architecture is explicitly organized around distinct zones—core infrastructure, learning management and support, monitoring, management, and control access—enabling modular design, clear separation of concerns, and smoother future extensions. In this way, the paper addresses a clear gap in the literature concerning CR environments dedicated to Wi‑Fi‑specific security experimentation and training, particularly for IEEE 802.11‑driven threat scenarios such as rogue access points, deauthentication attacks, and authentication‑level vulnerabilities. Part of the proposed architecture has already been realized as an open-source prototype~\cite{CRrepo}, covering the scenario generation, storage, retrieval, and instantiation workflow reflected in the functional and informational views. This implementation provides a practical foundation for future work toward a full‑scale realization of the platform, demonstrating that the conceptual architectural blocks can be instantiated in a namespace‑based IEEE 802.11 testbed built on \texttt{mac80211\_hwsim}, \texttt{hostapd}, \texttt{wpa\_supplicant}, and lightweight orchestration scripts. Overall, the work presented herein establishes a solid foundation for the development of Wi‑Fi‑specialized CRs, emphasizing targeted, reproducible, and educationally meaningful experimentation with IEEE 802.11 security scenarios.

\section*{Acknowledgment}

This work is supported by the Research Council of Norway through the SFI Norwegian Centre for Cybersecurity in Critical Sectors (NORCICS) project no. 310105 and by the European Union through the Horizon 2020 project PERSEUS (Grant No. 101034240).

\printbibliography

\end{document}